\documentclass[sigconf,screen]{acmart}

\usepackage{amssymb}
\usepackage{newtxmath}
\usepackage{bbm}

\AtBeginDocument{%
  }

\setcopyright{acmlicensed}
\copyrightyear{2025}
\acmYear{2025}
\acmDOI{XXXXXXX.XXXXXXX}
\acmConference[FSE '25]{ACM International Conference on the Foundations of Software Engineering}{June 23--27,
  2025}{Trondheim, Norway}
\acmISBN{978-1-4503-XXXX-X/18/06}

\begin{document}

\title{HyperSeq: A Hyper-Adaptive Representation for Predictive Sequencing of States}

\author{
    Roham Koohestani \quad Maliheh Izadi \\
    \{rkoohestani, m.izadi\}@tudelft.nl \\
    Delft University of Technology \\
    Delft, The Netherlands
}

\renewcommand{\shortauthors}{Koohestani et. al.}

\begin{abstract}
In the rapidly evolving world of software development, the surge in developers' reliance on AI-driven tools has transformed Integrated Development Environments into powerhouses of advanced features. This transformation, while boosting developers' productivity to unprecedented levels, comes with a catch: increased hardware demands for software development. Moreover, the significant economic and environmental toll of using these sophisticated models necessitates mechanisms that reduce unnecessary computational burdens. 
We propose HyperSeq — Hyper-Adaptive Representation for Predictive Sequencing of States — a novel, resource-efficient approach designed to model developers' cognitive states. HyperSeq facilitates precise action sequencing and enables real-time learning of user behavior. 
Our preliminary results show how HyperSeq excels in forecasting action sequences and achieves remarkable prediction accuracies that go beyond 70\%. Notably, the model's online-learning capability allows it to substantially enhance its predictive accuracy in a majority of cases and increases its capability in forecasting next user actions with sufficient iterations for adaptation. Ultimately, our objective is to harness these predictions to refine and elevate the user experience dynamically within the IDE.
\end{abstract}



\maketitle

\section{Introduction}
Since the advent of artificial intelligence, the software development life-cycle has become increasingly entangled with AI tooling.  
The introduction of tools like Copilot~\cite{copiot} and JetBrains AI~\cite{jetbrainsAI} is enhancing user productivity, while editors like Cursor~\cite{cursor} simplify programming, thereby transforming the way developers engage with software repositories.

Together with these benefits, it should be noted that the underlying models have added complexity to the systems we program, consequently raising the computational demands of programming. This increased computation demand has led to greater environmental repercussions; a recent technical blog by the team responsible for the Llama 3.1 model series reveals that training the leading 405 billion parameter model consumed 30.84 million GPU hours~\cite{llama31-2024}. This translates into 7.92 kilotons of CO2 emissions, based on the setup detailed in the article~\cite{grattafiori2024llama3herdmodels} and using the United States emission factor of 367 grams of $CO_2$ per kWh~\cite{us-energy-2024}. Furthermore, executing inferences with this model demands a multi-node arrangement, with each node equipped with 8 H100 GPUs operating at a 700W TDP, highlighting that a substantial impact of these models occurs during the post-deployment phase.

Previous studies have attempted to leverage user telemetry data to model user behavior~\cite{mozannar2024readinglinesmodelinguser} and to filter out completion requests in the IDE by taking into account the user's present state~\cite{mozannar2024show,de2024transformer}. Furthermore, earlier research has differentiated between two categories of user states: exploration, where a user seeks suggestions, and acceleration, where the user aims to continue programming without interruption.
However, these existing methods are still rather costly to train and require a large amount of training data. 

Moreover, customizing these models for individual users presents challenges, as it is quite improbable that a universal model could accurately capture the behavior of all users to the same satisfactory degree.
To tailor a model to an individual's behavior, online learning is required to consistently enhance the existing model data from that user, necessitating efficient implementation.

Hyperdimensional computing (HDC)\cite{kanerva2009hyperdimensional} provides a framework for holistic representation, manipulation, and querying of abstract concepts. This approach encodes ideas within high-dimensional vector spaces, taking advantage of the quasi-orthogonality inherent in these vectors.

This paper introduces HyperSeq, a Hyper-Adaptive Representation for Predictive Sequencing of States. We assess our model's effectiveness under two distinct scenarios, considering both adaptive and non-adaptive conditions using an established dataset of real-world user interactions and states. Our results indicate that the model effectively predicts the states of the users, with an average accuracy of more than 70\%. Moreover, leveraging adaptation mode allows the model to substantially improve its accuracy by employing online learning on user states.

In this paper, we (1) establish the theoretical basis for a hyperdimensional sequence model, (2) formally define predictive sequencing and demonstrate that the model allows for training in $O(D \cdot |Train|)$, and online adaptation in $O(D)$ time, (3) conduct an initial assessment of the model's performance, and (4) outline potential future applications for HyperSeq.

\section{Motivation}
There is a tremendous amount of cost associated with running these large models, especially when dealing with large-scale inference. 
An approach to addressing this issue involves minimizing the number of incorrect invocations, meaning decreasing the volume of calls that are likely to result in the rejection of generated responses. By analyzing user states, we can determine the likelihood of a user accepting or rejecting completions, thus alleviating server strain.

Moreover, recent research indicates~\cite{sergeyuk2024designspaceinidehumanai} that developers prefer to incorporate proactivity within their systems. By forecasting upcoming states and user actions, we can strive to embed state-aware mechanisms into IDEs. 

For example, detecting when a user is about to enter refactoring mode could allow the IDE to proactively initiate a search for potential refactorings and suggest these options as soon as the mode change is detected.

\section{Background Literature}
\subsection{Hyper-dimensional Computing}
Hyper-Dimensional Computing (HDC) is a computational paradigm in which high-dimensional vectors, typically ranging from $10^4$ to $10^5$, are utilized to represent, process and query abstract concepts. Although HDC was coined by Pentti Kanerva~\cite{kanerva2009hyperdimensional}, the concept dates to the 1990s, with holographic reduced representation (HRR)~\cite{plate1995holographic} and Vector-Symbolic Architectures~\cite{gayler2004vector}.

In recent years, there has been growing interest in employing HDC for intelligent tasks because of the efficiency of models based on HDC. Such HDC-based methods have been applied to challenging problems like Raven's Progressive Matrices~\cite{hersche2023neurovectorsymbolicarchitecturesolvingravens}, as well as in encoding visual representations efficiently and retrieving this data using what is referred to as resonator networks~\cite{kent2020resonatornetworksoutperformoptimization}.
Recent studies have explored the integration of HDC-inspired modifications into current architectures. InfiniAttention~\cite{munkhdalai2024leavecontextbehindefficient} has demonstrated significant advancements in processing exceptionally lengthy contexts by encoding them into a sparse hyperdimensional memory.

\subsection{User-IDE interaction}
Prior research has developed taxonomies of user interactions in IDEs, focusing on AI features. One study~\cite{barke2022groundedcopilotprogrammersinteract} identified two modes: exploration, where developers seek recommendations and solutions, and acceleration, where they work confidently with minimal distractions.

Various investigations have employed telemetry data from within the IDE to determine programmers' conditions. By analyzing features like caret movements and typing speed, these studies inferred the action states and developed a classification of 12 user states, along with their transitional characteristics~\cite{mozannar2024readinglinesmodelinguser}. The concept of leveraging telemetry data to enhance Human-AI interaction remains prevalent, with numerous studies applying it to effectively filter out requests that are highly likely to be declined~\cite{mozannar2024show,de2024transformer}.

Researchers have examined how to design Human-AI eXperiences (HAX) within the IDE by establishing a classification of methods through which developers seek assistance and identifying various domains where distinct models can provide support, including testing, maintenance, and optimization.

\subsection{Next-action prediction}
So far, limited research has focused on predicting users' next actions within IDEs. Cursor, a newly developed AI-integrated IDE, frequently references next-action prediction on its website, suggesting significant investment in this area~\cite{cursor}. Moreover, existing studies have explored employing gaze-tracking to dynamically adjust the IDE according to the user's forthcoming actions~\cite{10.1145/3603555.3603571}.

\section{Problem Definition}
In this section, we formally define the problem statement to improve clarity for subsequent sections. As noted in \autoref{sec:future_work}, our present work concentrates on predicting future user states given the limited availability of data. Nevertheless, we plan to broaden the investigation in future research to deduce subsequent user states directly from user telemetry.
\subsection{Predictive Sequencing}
Let \( S \) be the set containing all potential states, and let \( s \) indicate an individual state within \( S \). Similarly, let \( U \) signify the set of all users, with \( u \) representing a single user in \( U \). A session state sequence, denoted \( Sess \), is defined as an \( m \)-tuple of states 
\(
Sess = (s_0, s_1, \ldots, s_m), \text{where each } s_i \in S.
\)
Furthermore, we denote the \( i \)-th session of user \( u \) as \( Sess_{u,i} \). The objective of next-state prediction is to develop a model \( M \) capable of forecasting the following state for a user \( u \) based on an \( n \)-tuple of states. Formally, the model is articulated as:
\(
M: S^n \mapsto S.
\)
Additionally, we define a user-specific model as \( M_u \), tailored for the individual user \( u \).
\subsection{Hyper-Dimensional Computing (HDC)}
In this section, we expand upon the formal structure introduced earlier to lay the theoretical groundwork for our approach. We adhere to the following notational conventions:
\begin{itemize}
    \item The bundling operator is represented by \(\oplus\),
    \item The binding operator is signified by \(\otimes\), and 
    \item The permutation operator is symbolized by \(P\).
\end{itemize}

Further, we utilize the Bipolar Map framework, wherein hypervectors are drawn from the domain \( DOM = \{-1, 1\}^D \), with \( D \) denoting the dimensionality of the hyperspace. The elements of the chosen hypervectors are sampled from the probability distribution \( P = 2 \cdot (\text{Ber}(0.5) - 0.5) \)~\cite{gayler1998multiplicative}.

A hypervector \( v \) is characterized as \( v \in DOM \). A codebook \( C \) links the state space \( S \) to the hyperspace \( DOM \), i.e., \( C: S \mapsto DOM \). Likewise, \( C' \) provides the reverse mapping, defined as \( C': DOM \mapsto S \).
The function \( Encode \) is designed to process an \( n \)-tuple of hypervectors \( v_i \) while maintaining their sequential characteristics. Formally, \( Encode: DOM^m \mapsto DOM \), and it is implemented as following where \( V = (v_0, v_1, \ldots, v_{n-1}) \),
\[
Encode(V) = \bigotimes_{i=0}^{n-1}(P^{n-i-1}(v_i)).
\]

Additionally, we define a function \( Sim \) that takes two inputs from the defined hyperspace and yields a real-number indicating the similarity between the two vectors. More formally, \( Sim: (DOM, DOM) \mapsto \mathbb{R} \). In our context, we measure similarity using cosine similarity, with values ranging from -1 to 1, where -1 represents complete dissimilarity and 1 represents complete similarity.

\section{HyperSeq}
\subsection{Representation}
As stated earlier, the challenge of depicting action states involves encoding every action subsequence of length $n$ and then using the model to predict the most probable action from a subsequence of $n-1$ actions. Hence, a foundational model $M_{\text{Base}}$ can be developed utilizing the training dataset $Train$
\[
\bigoplus_{i=0}^{|Train| - 1}(Encode(Seq_i))
\]
where
\[
Train = \{Seq_0, Seq_1, \cdots, Seq_{|Train| - 1}\}
\]
\[
Seq_i = \{C(s_0), C(s_1), \cdots, C(s_{n})\}
\]
with $s_j$ being the ordered states within a session such that $Seq_i$ is a continuous subsequence of that session.
In the following sections, we explain the method used to derive the $Train$ and $Test$ datasets from our primary dataset $Data$.

The primary objective, naturally, of training a model of this kind is to utilize it for querying future user states, essentially engaging in predictive sequencing. This can be done by querying model $M$ with the prefix of the anticipated user-state. In other words, since the model is trained on n-grams of user states, the (n-1)-gram preceding the anticipated action can be used to query for the forthcoming action. The $Encode$ function allows for this (n-1)-gram prefix to be represented within the n-gram's context as a query vector \(q = P(Encode(Seq_k'[0:n-1]))\), where \(Seq_k'\) is intended to predict its n-th element.

By using our query vector $q$ with our Model $M$ through the binding operator $\otimes$, and due to the quasi orthogonality of the vectors, we solely unbind the suffix vector with a similar prefix (refer to ...). The resulting vector can be interpreted as a frequency vector representing the states linked to that prefix, expressed as $R = M * q = (C(s_0), C(s_1), \dots, C(s_l))$, where $l$ is determined by the occurrences of specific patterns within the training dataset. Because vector $R$ is a frequency vector, performing a similarity analysis of $R$ against all valid keys yields a discrete distribution pointing to the similarity with a specific state. A valid key $k$ is characterized by $k \in \text{Domain}(C')$, implying that the collection of all valid keys corresponds to the items for which mapping is available in the function $C'$, namely $\text{Domain}(C')$.

Thus, based on the frequency vector $R$, we can predict the n-th element in the sequence using $s_n = C'(\arg\max_{k}(Sim(R, k)))$. This methodology enables us to effectively perform predictive sequencing connected to action states.

\subsection{Adaptiveness}
Enhancing the foundation of the non-adaptive model, we propose altering our model's definition to enable greater flexibility. To accomplish this, we must revise the structure of our model $M$. This new iteration, model $M$, consists of two segments: $M_{\text{base}}$, akin to what was previously mentioned. Additionally, our model now incorporates an adaptive section, $M_{\text{adap}}$, which mirrors the base model's design. The complete model $M$ is formed by combining these two segments using the bundling $\oplus$ operator. The inclusion of this adaptive component supports online learning and facilitates incremental training, much like the base model. In the following section, we will examine how this adaptation influences model performance and evaluate the computational resources demanded by this additional component. The implementation details of our methodology, along with the evaluation procedure and results, are accessible in the paper's replication package\cite{replicationPackage}.

\section{Evaluation Methodology}
\subsection{Datasets}
Due to the challenges in accessing actual user data, we decided to utilize an already available dataset from a prior study, as outlined in the background section. Mozannar et al.~\cite{mozannar2024readinglinesmodelinguser} developed a taxonomy comprising twelve user states and gathered data from 21 developers to compile a labeled dataset of user states. This dataset can be employed to train and test the model in a constrained environment, allowing future opportunities for more comprehensive evaluation.
Our preliminary analysis reveals that three labels provide limited information, leading us to exclude them from this study. Consequently, we are left with a total of 9 labels.
\subsection{Data Partitioning Strategies}
Let \( Data \) denote the complete dataset of user session state sequences, where 
\(
Data = \{ Sess_{u,i} \mid u \in U', i \in \{1, \ldots, n_u\} \},
\)
\( U' \subset U \), and \( n_u \) is the number of sessions for user \( u \). The task is to split \( Data \) into two disjoint subsets, \( Train \) and \( Test \), such that \( Train \cup Test = Data \) and \( Train \cap Test = \emptyset \). We define three data partitioning strategies as follows.

\subsubsection{Disjoint Split}
In a disjoint split, the dataset is divided by users: one group for training and another for testing. We use data from 18 developers for training and the rest for testing.

\subsubsection{Overlapping Split}
In the overlapping split, each user's data is divided between training and testing sets. For example, with 10 sessions, 8 are for training and 2 for testing.

\subsubsection{K-Fold Cross-Validation}
This strategy creates a fold for each user by excluding them from training and using them only for evaluation, repeating for every user in the dataset.

\subsection{Hyper-Parameters}
In addition to exploring various data-splitting strategies, we investigate the impact of four other model (hyper-)parameters: dimension, subsequence length, cyclic shift, and adaptiveness. The dimension $D$ was adjusted, selecting from fixed values $\{1000, 5000, 10000, 20000\}$. Similarly, the sequence length was altered using the set $\{3, 5, 7, 9\}$. Moreover, within the MAP framework, where the $P$ operator acts as a cyclic shift of the representation vector, we assess the effect by altering the shift values among $\{2, 4, 6\}$. Lastly, we compare two proposed models: one incorporating the adaptive model and another without it.

\subsection{metrics}
\subsubsection{Overall Accuracy}
According to our setup, we assess the model's overall accuracy using the test set. When implementing cross-validation with several test sets, we calculate the average accuracy. For all other scenarios, we determine the model's accuracy considering every available user.

\subsubsection{Sliding Window Accuracy}
For adaptive models, we assess the sliding window accuracy to understand how the model's performance evolves. Specifically, a window of size $k$ is used to traverse the prediction sequence, and the model's accuracy is computed within this window. This approach helps us determine the effectiveness of employing online learning to tailor models to users.

\section{Preliminary Results}
\begin{figure}
    \centering
    \includegraphics[width=\linewidth]{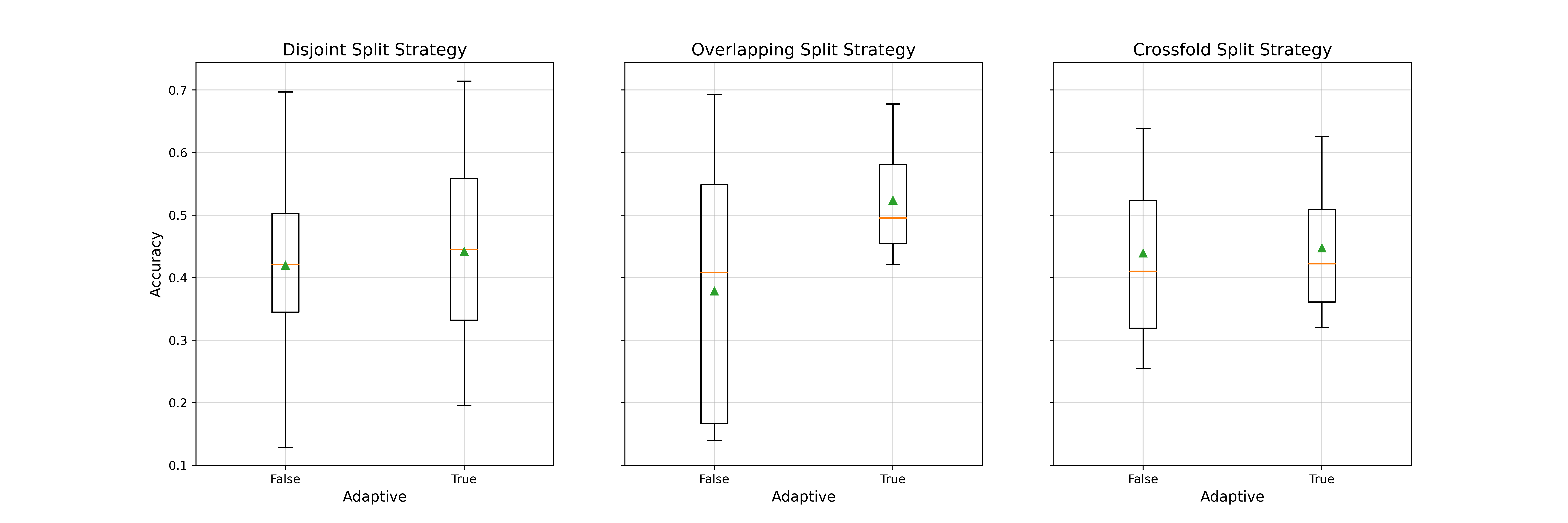}
    \caption{Overall accuracies of the model across the different splitting strategies and with adaptive on or off.}
    \label{fig:accuracies_avg}
\end{figure}
\begin{figure}
    \centering
    \includegraphics[width=\linewidth]{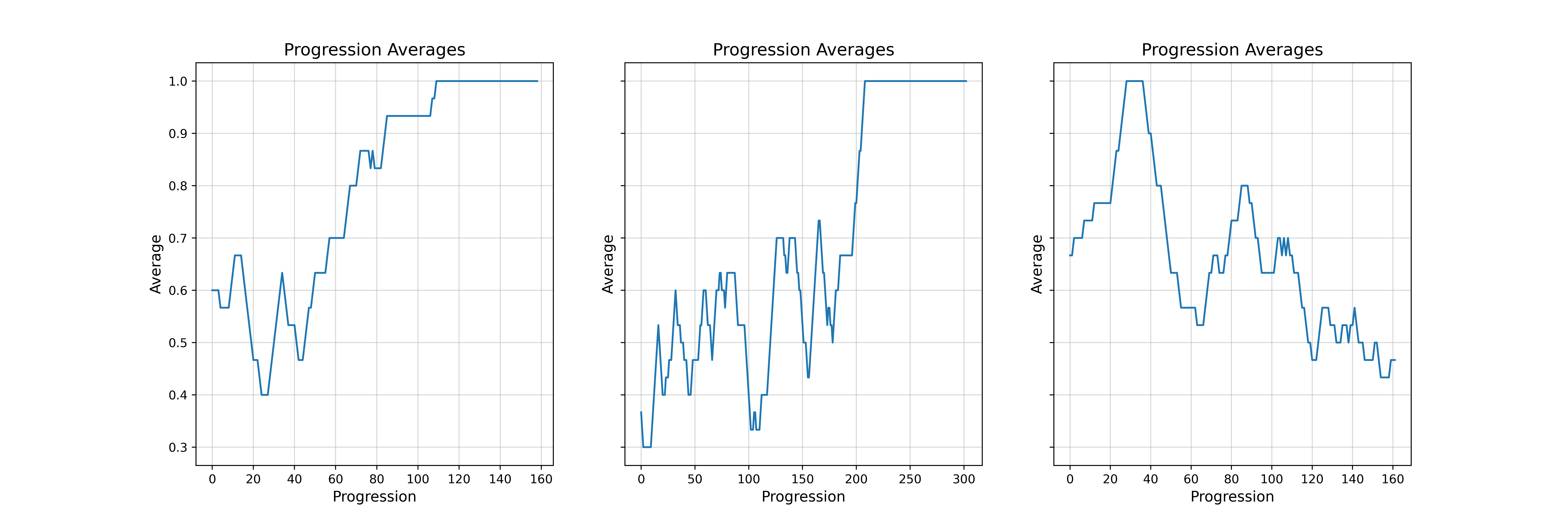}
    \caption{Best-performing model's accuracy; Sliding window size = 30,  Dimension = 20,000, Sequence Length = 3, Cyclic Shift = 4, Adaptive = True, and Split Strategy = Disjoint.}
    \label{fig:accuracies_average_progression}
\end{figure}
\subsection{Model Performance}
Our findings indicate that the models, on average, can perform within a range of 40 to 50 percent, with some configurations reaching above 70\% on average. Furthermore, as depicted in \autoref{fig:accuracies_avg}, the adaptive version of the model consistently surpasses the performance of the base version without employing any online learning.

Our assessments indicate that the cyclic shift parameter has a negligible impact on performance. Interestingly, our analysis of the results demonstrate that longer subsequences do not enhance predictive capabilities. In fact, an increase in sequence length corresponds with a decline in accuracy. This may result from the vector's information capacity being surpassed, suggesting further research is needed to determine if this pattern persists in higher dimensions.

Moreover, when examining the top-performing model, we observe a noteworthy trend in two of the three test scenarios: there is an upward trajectory in the model's average accuracy. This pattern suggests that online learning effectively enhances the model's performance. But more interestingly, there is not only an upward trend but rather that adapting the model can achieve results that make the predictions of the model near perfect.
\subsection{Model Efficiency}
\subsubsection{Training Efficiency}
Training the model involves a single pass through the training data. N-grams can be constructed in constant time, provided that previous sequences within a session are available and that the cold start of the session is amortized (see the appendix). Since the model operates on D-dimensional vectors, the time complexity for training is $O(D \cdot |Train|)$.

\subsubsection{Update Efficiency}
The update logic is implemented as a constant amount of operations on D dimensional vector, making the update $O(D)$.

\subsubsection{Storage Efficiency}
The storage requirements for the model depend on several components, which vary based on the specific implementation of the approach. To optimize memory usage, the domain of stored values per entry can be limited, allowing for storage in fewer bits than the conventional 32 or 64 bits. For instance, using 8 or 16 bits per entry can enhance storage efficiency. Let B represent the number of bits per entry. The stored items include the model memory (doubled in the case of the adaptive model) and the codebook. Consequently, the space complexity of the model is $O(B \cdot (|C| + D))$.

\section{Future Work}
\label{sec:future_work}
In future work, we intend to broaden our approach for more extensive, real-world applicability. We plan to enhance the proposed framework to process incoming telemetry data from users similarly to other current methodologies. Moreover, we seek to leverage the predicted user state to adaptively optimize the user experience within the IDE. Furthermore, we are interested in investigating the impact of different Vector-Symbolic Architectures and assessing how they influence model accuracy.

\section{Conclusion}
This paper introduced HyperSeq, a Hyper-Adaptive Representation for Predictive Sequencing of States, as a novel, efficient approach to predicting developer actions in IDEs. Using hyperdimensional computing and online adaptation, HyperSeq achieved competitive accuracy, surpassing 70\% in certain configurations, while maintaining low computational and storage requirements. Our evaluation demonstrated that HyperSeq not only excels in forecasting user actions but also dynamically improves its accuracy through online learning, enabling personalized predictions for individual users. This efficiency positions HyperSeq as a promising solution to reduce computational overhead in AI-powered IDEs and enhance user experiences through context-aware proactive adaptations. Future work will focus on integrating telemetry-based predictions, exploring alternative vector-symbolic architectures, and deploying HyperSeq in real-world environments to optimize developer productivity and promote sustainable computing practices.

\bibliographystyle{ACM-Reference-Format}
\bibliography{main}
\clearpage
\appendix
\section{The result of querying the model M with q is a frequency vector R}
We define a \emph{frequency vector} as a vector \(F\) in hyperspace that represents the weighted combination of a set of basis vectors \(s_0, s_1, \dots, s_m\), such that:
\[
R = a \cdot f_0 \oplus b \cdot f_1 \oplus \cdots \oplus z \cdot f_m,
\]
where \(a, b, \dots, z \geq 0\) and each weight corresponds to the frequency or relevance of the respective basis vector.

The model is trained as:
\[
M = \bigoplus_{i=0}^{|Train| - 1}(Encode(Seq_i))
\]
where:
\[
Train = \{Seq_0, Seq_1, \dots, Seq_{|Train| - 1}\},
\]
and each sequence is defined as:
\[
Seq_i = \{C(s_0), C(s_1), \dots, C(s_n)\}.
\]

By the definition of the \(Encode\) function, the model formulation can be expanded as:
\[
M = \bigoplus_{i=0}^{|Train| - 1}\left(P(Encode(Seq_i[:n-1])) \otimes Seq_{i,n}\right).
\]

For brevity, we define \(Seq_i[n-1]\) as \(Seq_{i,n}\). Similarly, the query vector \(q\) is defined as:
\[
q = P(Encode(Seq_k'[0:n-1])).
\]

Due to the high quasi-orthogonality and the low probability of collisions between vectors in hyperspace, multiplicative unbinding (or querying, as described in the paper) results in the expression:
\[
M \otimes q = \bigoplus_{i=0}^{|Train| - 1}\left(q \otimes P(Encode(Seq_i[:n-1])) \otimes Seq_{i,n}\right).
\]

This simplifies further under the assumption that all subsequences \(Seq_i[:n-1]\) in the training data are unique and match the queried subsequence. In this case, the formulation becomes:
\[
M \otimes q = \bigoplus_{i=0}^{|Train| - 1}(\mathbbm{I} \otimes Seq_{i,n}),
\]
where \(\mathbbm{I}\) is the identity vector. 

However, this assumption is unlikely in practice. When the preceding subsequence does not match exactly, the quasi-orthogonality ensures that the similarity between unrelated vectors is normally distributed around zero, resulting in negligible noise. Thus, in the absence of matching subsequences, \(M \otimes q\) simplifies to noise, which is equally dissimilar to all vectors.

Therefore, in the general case, \(M \otimes q\) can be expressed as:
\[
\bigoplus_{i=0}^{|Train| - 1}(Seq_{i,n}) + noise.
\]

This represents a frequency vector of possible states. The resulting vector \(R\) can be written as:
\[
R = a \cdot s_0 \oplus b \cdot s_1 \oplus \cdots \oplus z \cdot s_m,
\]
where \(a, b, \dots, z \geq 0\).

This is by definition a frequency vector and by the nature of the hyperspace, the vector \(R\) will be most similar to \(s_j\), where the corresponding factor (e.g., \(a, b\)) is the greatest.

\section{N-grams can be constructed in amortized constant time}
\textbf{Claim:} 
\emph{Let $\bigl(s_0, s_1, \dots, s_{m-1}\bigr)$ be a session of length $m$ from which we form $n$-grams. Once the first $n$-gram is built (cold start) and amortized, constructing each subsequent $n$-gram (i.e., sliding the window by one) requires only constant time, irrespective of $m$.}

Let $s_i$ denote the state at index $i$ within a session of length $m$. We define an $n$-gram at index $i$ by $\bigl(s_i,\, s_{i+1},\, \dots,\, s_{i+n-1}\bigr)$. Let $G_i$ denote the hypervector encoding of the $n$-gram starting at $i$:
\[
    G_i = \mathrm{Encode}\bigl(s_i, s_{i+1}, \dots, s_{i+n-1}\bigr).
\]

\begin{theorem}
\label{thm:constant-time}
Suppose we have a session of length $m$ and we wish to form every $n$-gram $G_i$ for $i=0,1,\dots,m-n$. Once $G_0$ is computed and its cost amortized, $G_{i+1}$ can be computed from $G_i$ in constant time. Consequently, forming all $n$-grams in the session costs $\mathcal{O}(m)$ (excluding operations in dimension $D$), i.e., constant time per $n$-gram.
\end{theorem}

\subsection{Proof}

\paragraph{1. Amortized Cold-Start.}
The first $n$-gram $G_0$ is computed by encoding $\bigl(s_0, s_1, \dots, s_{n-1}\bigr)$.
This cold-start takes $\mathcal{O}(n)$ work plus dimensional operations $\mathcal{O}(D)$ (where $D$ is the dimension of the hypervectors). We spread this cost over all $m-n+1$ $n$-grams in the session. Thus, amortized per $n$-gram, the cold-start overhead becomes negligible as $m\gg n$. Formally, $\frac{n + \mathcal{O}(D)}{m-n+1} \approx \mathcal{O}\!\bigl(\tfrac{1}{m}\bigr)$ when $m\gg n$.

\paragraph{2. Sliding Window Update: $G_i \to G_{i+1}$.}
Assume $G_i$ is known. We need $G_{i+1} = \mathrm{Encode}(s_{i+1}, \dots, s_{i+n})$. Under typical vector-symbolic architectures:
\[
    G_{i+1}
    \;=\;
    \mathrm{Update}\bigl(G_i,\; s_i,\; s_{i+n}\bigr),
\]
where $Update$ ``removes'' the effect of $s_i$ and ``adds'' the effect of $s_{i+n}$ via hyperdimensional operations. For instance, if $\mathrm{Encode}$ used a binding operator $\otimes$ that behaves like elementwise multiplication, then removing a state amounts to multiplying by the inverse of its hypervector (here, the hypervector inverse is the same as the hypervector itself in $\{-1,+1\}^D$). Adding a new state is another elementwise multiply or combine step.

Crucially, this sliding update from $G_i$ to $G_{i+1}$ involves a \emph{constant} amount of small, discrete operations (unbinding one vector, binding another). The \emph{number} of operations does not scale with $i$ or $m$. Thus,
\[
    G_{i+1} \longleftarrow G_i 
    \quad\text{in }\mathcal{O}(1)\text{ (excluding dimension }D\text{)}.
\]

\paragraph{3. Per-Step Cost is Constant.}
For each new position $i$, we spend $\mathcal{O}(1)$ time to compute $G_{i+1}$ from $G_i$, apart from the $\mathcal{O}(D)$ dimensional operations that do not depend on $m$. Therefore, constructing all $n$-grams in a session of length $m$ involves $(m-n+1)$ incremental steps, each in $\mathcal{O}(1)$ plus dimension $D$ factors.

\paragraph{4. Total Complexity.}
Over one session, forming every $n$-gram therefore costs:
\[
    \mathcal{O}(n + (m-n)\times 1) = \mathcal{O}(m),
\]
plus the dimensional factor $\mathcal{O}(D)$ per step. Importantly, it is \emph{independent of $m$} in the sense that each step is constant. When summing over a large training set $|\mathrm{Train}|$, this procedure extends naturally to yield $\mathcal{O}\!\bigl(D\times|\mathrm{Train}|\bigr)$ for the entire dataset in a single pass.

$\blacksquare$

\end{document}